\documentclass[preprint,amsmath,amssymb,aps,showkeys,showpacs,pra]{revtex4-1}

\usepackage[english]{babel}
\usepackage{graphicx}
\usepackage{graphics}
\usepackage{amsmath}
\usepackage{dcolumn}
\usepackage{amssymb}
\usepackage{bm}
\usepackage[latin1]{inputenc}
\usepackage{graphicx,color}
\usepackage{subcaption}

\begin{document}
\title{Landau levels for massive disclinated graphene-based topological insulator}

\author{G. Q. Garcia}
\email{gqgarcia99@gmail.com}
\affiliation{Departamento de F\'isica, Universidade Federal da Para\'iba, Caixa Postal 5008, 58051-970, Jo\~ao Pessoa, PB, Brazil.} 

\author{J. R. S. Oliveira}
\email{jardson.ricardo@gmail.com}
\affiliation{Departamento de F\'isica, Universidade Federal da Para\'iba, Caixa Postal 5008, 58051-970, Jo\~ao Pessoa, PB, Brazil.} 

\author{P.J. Porf\'irio}
\email{pporfirio@fisica.ufpb.br}
\affiliation{Departamento de F\'isica, Universidade Federal da Para\'iba, Caixa Postal 5008, 58051-970, Jo\~ao Pessoa, PB, Brazil.}

\author{C. Furtado}
\email{furtado@fisica.ufpb.br} 
\affiliation{Departamento de F\'isica, Universidade Federal da Para\'iba, Caixa Postal 5008, 58051-970, Jo\~ao Pessoa, PB, Brazil.} 

\begin{abstract}
In this work, we investigate the massive Kane-Mele model for graphene in the presence of disclination and an external magnetic field, where graphene behaviors as a topological insulator. In the low-energy limit, the effective field equation for graphene is described by a Dirac equation with three different degrees of freedom. We succeed to decouple the set of eight components of the Dirac equation by using the spin projector $\hat{C}$ in the disclinated geometry. As consequence, we obtain the Landau levels in this framework, in which we note the emergence of zero modes as edge states due to the inversion symmetry breaking. We also note that for different sites in sublattices $\mathcal{A/B}$, one can have different values of gap width. 
\end{abstract}
\maketitle
\section{Introduction}

Following the results that the quantum Hall effect was obtained by confining electrons in two dimensions and subjecting them to an external magnetic field \cite{Von}, a new kind of material called a Topological Insulator (TI) was introduced. The topological insulators is characterized by its bulk being an insulator, while it presents metal edge states that are topologically protected \cite{Shen, Moore}. Indeed, TIs represent a different phase of matter where their properties are protected from small changes in the material, such that only a phase transition can alter them \cite{Hasan, Fruchart}. The preservation or lack thereof of the symmetries present in a lattice of a certain material is of fundamental importance for the classification of topological insulators \cite{Ando}. Thus, the inversion and time-reversal symmetries have been broadly studied for different kind of materials, in order to discovery a variety of TIs. In 2004, Kato {\it et al.} observed the spin Hall effect for the strained gallium arsenide \cite{Kato}. Kane and Mele introduced a $Z_2$ topological invariant for classification of TIs \cite{Kane}. After that, Fu and Kane used this $Z_2$ topological invariant to predict a variety of materials as topological insulators \cite{Fu}, among them are the semiconducting alloy $Bi_{1-x} Sb_{x}$ and the $\alpha-Sn$ under a uniaxial strain. In ref. \cite{Roy}, Roy demonstrated that the edge states is related with the topological invariant formulation. From the effective topological field theory, Qi {\it et. al} showed the appearance of a magneto-electric effect on the surface states for a two band model \cite{Qi}. 

Since the discovery of graphene by Novoselov and Geim \cite{Novoselov, Geim}, a graphene layer was immediately pointed out as a candidate to be a topological insulator (TI). Graphene has a two-dimensional structure, where its lattice is composed of carbon atoms arranged in a honeycomb pattern. Due to its symmetry properties, including inversion and time-reversal symmetries, graphene exhibits topologically protected zero modes \cite{Katsnelson}. Even before this discovery, studies involving the honeycomb lattice already indicated the possibility of this type of material being a TI. In 1988, Haldane described a honeycomb lattice that breaks these two symmetries: inversion and time-reversal symmetries \cite{Haldane}. As a consequence of the symmetry breaking, Haldane obtained a topological insulator even without a magnetic field, characterized by a non-zero Chern number. Following this, Kane and Mele studied the effects of spin-orbit interactions in graphene \cite{Kane2}. They reintroduced time-reversal symmetry in the Haldane model by incorporating spin degrees of freedom with Rashba-type spin-orbit coupling (SOC) contributions. Consequently, graphene was converted into a different type of topological insulator. An important characteristic of graphene is the possibility to study topological defects through the geometric theory of defects by Katanev and Volovich \cite{Katanaev}. Building on this, Choudhari and Deo used this theory to introduce disclinations in the Kane-Mele model for graphene, demonstrating that this topological defect can represent a type of Rashba spin-orbit coupling \cite{Choudhari}. Recently, we are exploring the graphene as topological insulator candidate with works series \cite{Garcia1, Garcia2}. In particular, we have showed the rise of Landau levels for the graphene-based topological insulator in different context. 

In this paper, we aim to study the topological insulator based on graphene. For this purpose, we describe the massive graphene using the Kane-Mele model with a disclination to simulate a Rashba spin-orbit term. Additionally, we introduce an external magnetic field perpendicular to the graphene sheet. Our objective is to obtain the Landau levels for our graphene-based topological insulator. In Sec. \ref{sec2}, we describe the Kane-Mele Hamiltonian with disclination in the presence of an external magnetic field. Subsequently, in Sec. \ref{sec3}, we present the method to solve this Hamiltonian. Finally, in Sec. \ref{sec4}, we derive the Landau levels for the graphene-based topological insulator and analyze the energy levels and the eight-component spinor. In the concluding Sec. \ref{sec5}, we discuss our final considerations.

\section{Massive Kane-Mele model for graphene with disclination}\label{sec2}

Let us begin with the description of a graphene layer considering the spin-orbit coupling contribution. For this purpose, we describe a graphene layer using the Kane-Mele model \cite{Kane2}. In the tight-binding model for graphene, we consider two contributions. The first one is related to electron hopping to its nearest neighbors, i.e., an electron in sublattice $\mathcal{A/B}$ has the probability to hop to its nearest neighbor in sublattice $\mathcal{B/A}$, respectively. The second one involves spin-orbit coupling (SOC), where we introduce a next-nearest neighbors hopping term. Additionally, we introduce an extra term representing an effective mass for graphene. Putting all these ingredients together, we end up with the massive Kane-Mele Hamiltonian:
\begin{eqnarray}
    H = t\sum_{\langle ij\rangle\alpha} c^{\dagger}_{i\alpha} c_{j\alpha} + \sum_{\langle\langle ij\rangle\rangle\alpha\beta} it_{2}\nu_{ij} s^{z}_{\alpha\beta} c^{\dagger}_{i\alpha} c_{j\beta} + M c^{\dagger}_{i\alpha} c_{i\alpha}.
    \label{2.1}
\end{eqnarray}
Note that the SOC term is responsible for the preservation of time-reversal symmetry and introduces spin dependence in electron transitions among next-nearest neighbors. The massive term breaks inversion symmetry. Here, the parameter $\nu_{ij} = \pm 1$ depends on the orientation of the second-neighbor hopping, and the operator $s^{z}_{\alpha\beta}$ is the spin projector in the $z$-direction. In the low-energy limit, the Kane-Mele model (\ref{2.1}) is given by
\begin{eqnarray}
    H(\vec{k}) = \hslash v_f (\tau_z \sigma_x k_x + \sigma_y k_y) + \lambda_{SO} \tau_z \sigma_z s_z + M\tau_z\sigma_z,
    \label{2.2}
\end{eqnarray}
where the parameter $v_f$ is the Fermi velocity and $\lambda_{SO} = 3\sqrt{3}t_2$ is the intrinsic spin-orbit coupling parameter. The Kane-Mele Hamiltonian presents three sets of Pauli matrices, each of which corresponds to different degrees of freedom. The Pauli matrices $\tau_i$ are related to the valley degrees of freedom at $K$ and $K'$. The second set of Pauli matrices, $\sigma_i$, corresponds to the sublattices $\mathcal{A/B}$. The spin degrees of freedom are represented by the Pauli matrices $s_i$. The Kane-Mele Hamiltonian is an $8 \times 8$ matrix acting on an eight-component spinor. 

In order to build a graphene layer with disclination, we shall make use of the Volterra processes \cite{Volterra, Puntingan}, which consist of the removal or insertion of angular graphene sectors in a flat structure to explain this kind of defect. For the disclination case, this angular sector is given by $\lambda = N\frac{\pi}{3}$. For $N = -1$, an angular sector is inserted, resulting in a heptagon ring at the defect core. For $N = 1$, an angular sector is removed, resulting in a pentagon ring at the cone apex. In the low-energy limit, we can use the geometric theory of defects \cite{Katanaev} to describe topological defects by the metric tensor. For the disclination geometry, the corresponding line element is write as,
\begin{eqnarray}
    ds^2 = dt^2 - dr^2 - \alpha^{2}_{N} r^2 d\theta^{2},
    \label{2.3}
\end{eqnarray}
where $\alpha_{N} = 1 - \frac{N}{6}$ and $0<r<\infty$ and $0<\theta<2\pi$. When we perform a parallel transport of the spinor around the defect, it provides a variation of the local reference frame, resulting in gauge transformations to compensate for the phase differences among the spinor components \cite{Choudhari}. This general gauge transformation is presented by holonomy
\begin{eqnarray}
    \psi\left(\phi = 2\pi - \frac{n\pi}{3}\right) = e^{i\frac{n\pi}{6}(\tau_z \sigma_z s_0 - 3\tau_y \sigma_y s_0)}\psi(\phi = 0),
    \label{2.4}
\end{eqnarray}
where $\phi = \frac{\theta}{\alpha_N}$ is the angular coordinate for the unfolded plane. Note that the holonomy presented in equation (\ref{2.4}) has two contributions: The first one, $U_s(\phi) = e^{i\frac{\phi}{2}\tau_z\sigma_z s_0}$, is responsible for the spinor co-rotation transformation, and the second one, $V_{Ns}(\theta) = e^{i\frac{N\theta}{4}\tau_y\sigma_y s_0}$, represents a kind of Aharonov-Bohm quantum flux. Consequently, a quantum flux appears at the center of the defect. We can introduce this quantum flux via the vector potential
\begin{eqnarray}
    \vec{A}_{\Phi} = \frac{\Phi}{r\alpha_N\Phi_0}\hat{\theta},
    \label{2.5}
\end{eqnarray}
where $\Phi_0 = h/2e$ is the magnetic quantum flux. Besides this holonomy, we are interested in how the external magnetic field acts on this topological insulator. Therefore, let us introduce a constant magnetic field $\vec{B} = B_0 \hat{z}$ through
\begin{eqnarray}
    \vec{A} = \frac{B_0 r}{2}\hat{\theta}.
    \label{2.6}
\end{eqnarray}

 By taking into account the holonomy (\ref{2.4}) and both potential vectors (\ref{2.5}) and (\ref{2.6}) and substituting them into the massive Kane-Mele Hamiltonian (\ref{2.2}), it results in
\begin{eqnarray}
    H = \left(k_r -\frac{i}{2r}\right)\tau_z\sigma_x + \left(k_\theta + \frac{\Phi}{r\alpha_N\Phi_0} \pm \frac{N}{4r\alpha_N} + \frac{eB_0}{2}r\right)\sigma_y + M\tau_z\sigma_z + H_{SO},
    \label{2.7}
\end{eqnarray}
where the spin-orbit Hamiltonian $H_{SO}$ in the disclinated background (\ref{2.3}) is,
\begin{eqnarray}
    H_{SO} = \lambda_{SO}\left(h(\beta)\tau_z\sigma_z s_z + p(\beta)\tau_z\sigma_z s_x\right),
    \label{2.8}
\end{eqnarray}
with $p(\beta) = \frac{2\cos{\beta}}{1+\sin{\beta}}$ and $h(\beta) = 1 -\frac{p^2(\beta)}{4}$, where $\beta$ is the conical aperture angle. The spin-orbit coupling term obtained in equation (\ref{2.8}) is a Rashba-type term. The massive Kane-Mele Hamiltonian acts on the eight component spinor,
\begin{eqnarray}
    \Psi = \left( (\psi^{K}_{A\uparrow}\ \psi^{K}_{A\downarrow}\ \psi^{K}_{B\uparrow}\ \psi^{K}_{B\downarrow}) (\psi^{K'}_{A\uparrow}\ \psi^{K'}_{A\downarrow}\ \psi^{K'}_{B\uparrow}\ \psi^{K'}_{B\downarrow})\right)^{T}.
    \label{2.9}
\end{eqnarray}
\section{Solutions for massive Kane-Mele Hamiltonian}\label{sec3}

Our goal is to demonstrate the existence of Landau levels for the Kane-Mele Hamiltonian with disclination and in the presence of an external magnetic field. To achieve this, it is necessary to solve the Schr\"odinger equation for the massive Kane-Mele Hamiltonian (\ref{2.7}). Let us rewrite the Hamiltonian $H$ as follows:
\begin{eqnarray}
    \left[\hat{\pi}_r\tau_z \sigma_x + \hat{\pi}_\theta\sigma_y + M\tau_z \sigma_z + \lambda_{SO} \left(h(\beta)\tau_z\sigma_z s_z + p(\beta) \tau_z \sigma_z s_x\right)\right]\Psi = E\Psi,
    \label{3.1}
\end{eqnarray}
where we define the conjugate momentum operators as
\begin{eqnarray}
    \hat{\pi}_r = \left[k_r - \frac{i}{2r}\right],\quad \hat{\pi}_\theta = \left[k_\theta + \frac{\Phi}{r\alpha_N \Phi_0} \pm \frac{N}{4r\alpha_N} + \frac{eB_0}{2}r\right].
    \label{3.2}
\end{eqnarray}
Here, the momentum operators are given by $k_r = -i\partial_r$ and $k_\theta = -i\partial_\theta/r\alpha_N$. As previously mentioned, the Kane-Mele Hamiltonian is an $8\times 8$ matrix, divided into two blocks representing the valleys $K$ and $K'$. Therefore, we can express the Hamiltonian operator as
\begin{eqnarray}
    H = \left(\begin{array}{cc}
        H_K & 0 \\
        0 & H_{K'}
    \end{array}\right).
    \label{3.3}
\end{eqnarray}
Each matrix element on the main diagonal is a $4\times 4$ matrix, corresponding to the $\mathcal{A/B}$ sublattices and the spin degrees of freedom. In other words, we have two independent Dirac equations for the valleys $K$ and $K'$.

In order to solve these two Dirac equations, we need to rewrite both in terms of a different representation of Dirac matrices. In this case, we can express the Dirac equations as follows:
\begin{eqnarray}
    H_K &= \gamma^1\hat{\pi}_r + \gamma^2 \hat{\pi}_\theta + \gamma^5\gamma^0\gamma^3\left( M + \gamma^0\lambda_{SO}\left[\gamma^3 p(\beta) - \gamma^5 h(\beta)\right]\right), \label{3.4}\\
    H_{K'} &= -\gamma^1\hat{\pi}_r + \gamma^2 \hat{\pi}_\theta - \gamma^5\gamma^0\gamma^3 \left( M + \gamma^0\lambda_{SO}\left[\gamma^3 p(\beta) - \gamma^5 h(\beta)\right]\right). \label{3.5}
\end{eqnarray}
We built a different representation of Dirac matrices which obeys the Clifford algebra $\lbrace\gamma^{\mu},\gamma^{\nu}\rbrace = 2\eta^{\mu\nu} I_{4\times 4}$, and they are defined by
{\small
\begin{eqnarray}
    \gamma^{0} &= \left(\begin{array}{cc}
      i\sigma^2 & 0 \\
        0 & -i\sigma^2
    \end{array}\right),\ 
     \gamma^{1} = \left(\begin{array}{cc}
      0 & I_{2\times 2} \\
        I_{2\times 2} & 0
    \end{array}\right),\ 
     \gamma^{2} = \left(\begin{array}{cc}
      0 & -iI_{2\times 2} \\
        iI_{2\times 2} & 0
    \end{array}\right),\ 
     \gamma^{3} = \left(\begin{array}{cc}
      \sigma^3 & 0 \\
        0 & -\sigma^3
    \end{array}\right). \label{3.6}
\end{eqnarray}}
The fifth gamma matrix is defined as $\gamma^5 = i\gamma^0\gamma^1\gamma^2\gamma^3$, so that $\lbrace\gamma^5,\gamma^\mu\rbrace = 0$ and $(\gamma^5)^2 = I_{4\times 4}$. Therefore, we can write the fifth gamma matrix $\gamma^5$ as
\begin{eqnarray}
    \gamma^5 = \left(\begin{array}{cc}
    \sigma^1 & 0 \\
       0 & -\sigma^1
    \end{array}\right). \label{3.7}
\end{eqnarray}
From these definitions, we can finally express the Schr\"odinger equation (\ref{3.1}) as
{\footnotesize
\begin{eqnarray}
    \left(\begin{array}{cc}
    \gamma^1\hat{\pi}_r + \gamma^2 \hat{\pi}_\theta + \Sigma^3\left( M + \hat{C}\right) & 0 \\
    0 & -\gamma^1\hat{\pi}_r + \gamma^2 \hat{\pi}_\theta - \Sigma^3\left( M + \hat{C}\right)
    \end{array}\right) \left(\begin{array}{c}
    \psi^{K} \\
    \psi^{K'}
    \end{array}\right) = E \left(\begin{array}{c}
    \psi^{K} \\
    \psi^{K'}
    \end{array}\right), \label{3.8}
\end{eqnarray}}
where we have introduced the spin matrices $\Sigma^i = \gamma^5\gamma^0\gamma^i$ for $i=1,2,3$ in this new Dirac matrix representation. The operator $\hat{C}$ is defined by
\begin{eqnarray}
    \hat{C} = \lambda_{SO}\gamma^0\left(\gamma^3 p(\beta) - \gamma^5 h(\beta)\right). \label{3.9}
\end{eqnarray}

It is important to observe that the equations (\ref{3.4}) and (\ref{3.5}) do not depend on the $\theta$-coordinate, and therefore $\hat{\pi}_\theta$ commutes with both Hamiltonian operators. Additionally, the $\hat{C}$ operator is also a constant of motion, meaning that the commutator between this operator and the Hamiltonian operators $[H_K,\hat{C}] = [H_{K'},\hat{C}]$ is zero. As a consequence, the $\hat{C}$ operator shares its basis with these two Hamiltonian operators for the valleys $K$ and $K'$. The eigenvalue equation for the $\hat{C}$ operator
\begin{eqnarray}
    \hat{C}\Psi = \nu\lambda_{SO}\sqrt{h^{2}(\beta) + p^2(\beta)}\Psi,
    \label{3.10}
\end{eqnarray}
provides the eigenvalues for the $\hat{C}$ operator, with the parameter $\nu = \pm 1$. It is also possible to obtain the shared basis between $\hat{C}$ and $H$. Thus, we can write the solution for the Schr\"odinger equation (\ref{3.8}) as
\begin{eqnarray}
    \Psi(r,\theta) = e^{ij\theta} \left(\begin{array}{c}
    \psi^{K}(r)\\
    \psi^{K'}(r)
    \end{array}\right), \label{3.11}
\end{eqnarray}
where $j = \pm\frac{1}{2}, \pm \frac{3}{2}, \ldots$ is the angular momentum eigenvalue. The wavefunctions $\psi^{K}$ and $\psi^{K'}$ are given by
\begin{eqnarray}
    \psi^{V} = \left(\begin{array}{c}
    \psi^{V}_{A} \\ 
    c_{\nu}\psi^{V}_{A} \\ 
    \psi^{V}_{B} \\ 
    c_{\nu}\psi^{V}_{B} 
    \end{array}\right), \label{3.12}
\end{eqnarray}
where $V$ represents the valleys $K$ and $K'$, and
\begin{eqnarray}
    c_\nu = - \frac{h(\beta) - \nu\sqrt{h^{2}(\beta) + p^{2}(\beta)}}{p(\beta)}. \label{3.13}
\end{eqnarray}
It is essential to recognize that the operator $\hat{C}$ acts as the spin projector in the disclinated geometry.

\section{Landau levels in massive Kane-Mele Hamiltonian}\label{sec4}

Having developed the main mathematical tools in the previous sections, we are now able to solve both independent Dirac equations, and thus to obtain the Landau levels for a Kane-Mele Hamiltonian with disclination in the presence of an external magnetic field. Taking the solution (\ref{3.11}) and substituting it into the Schr\"odinger equation (\ref{3.8}), we obtain two coupled differential equations for each valley. For the $K$ valley, we have
\begin{eqnarray}
    -i\left[\left(\frac{d}{dr} + \frac{1}{2r}\right) - \frac{1}{r}\left(\mu_{\tau} + \frac{eB_0}{2}r^2\right)\right]\psi^{K}_{A} = \left[E + \left(M + \nu\lambda_{SO}\sqrt{h^2 + p^2}\right)\right]\psi^{K}_{B};
    \label{4.1}\\
    -i\left[\left(\frac{d}{dr} + \frac{1}{2r}\right) + \frac{1}{r}\left(\mu_{\tau} + \frac{eB_0}{2}r^2\right)\right]\psi^{K}_{B} = \left[E - \left(M + \nu\lambda_{SO}\sqrt{h^2 + p^2}\right)\right]\psi^{K}_{A},
    \label{4.2}
\end{eqnarray}
and for the $K'$ valley, we also have
\begin{eqnarray}
    i\left[\left(\frac{d}{dr} + \frac{1}{2r}\right) + \frac{1}{r}\left(\mu_{\tau} + \frac{eB_0}{2}r^2\right)\right]\psi^{K'}_{A} = \left[E - \left(M + \nu\lambda_{SO}\sqrt{h^2 + p^2}\right)\right]\psi^{K'}_{B};
    \label{4.3}\\
    i\left[\left(\frac{d}{dr} + \frac{1}{2r}\right) - \frac{1}{r}\left(\mu_{\tau} + \frac{eB_0}{2}r^2\right)\right]\psi^{K'}_{B} = \left[E + \left(M + \nu\lambda_{SO}\sqrt{h^2 + p^2}\right)\right]\psi^{K'}_{A},
    \label{4.4}
\end{eqnarray}
where we have used the shorthand notation:
\begin{eqnarray}
    \mu^{V} = \frac{j+\frac{\Phi}{\Phi_0} + \tau_z\frac{N}{4}}{\alpha_N},
    \label{4.5}
\end{eqnarray}
where $\tau_z = \pm 1$, in order to express the eigenvalues for the $\hat{\pi}_\theta$ conjugated momentum. To decouple both sets of first-order differential equations, we substitute one differential equation into another from the same set. Thus, we obtain two decoupled differential equations for each valley. In a compact form, the differential equation has the following way,
\begin{eqnarray}
    \frac{d^2 \psi^{V}_{i}}{dr^2} + \frac{1}{r}\frac{d\psi^{V}_{i}}{dr} &-& \left[ \frac{1}{r^2}\left(\mu^V \mp \frac{\sigma_z}{2}\right)^2 \right. + \frac{e^2 B_0^2}{4}r^2 +\nonumber \\
    &+& \left. eB_0 \left(\mu^V \pm \frac{\sigma_z}{2}\right) + \left(M + \nu\lambda_{SO}\sqrt{h^2 + p^2}\right)^2 - E^2 \right]\psi^{V}_{i} = 0.
    \label{4.6}
\end{eqnarray}
Here, $\sigma_z = \pm 1$. This implies that we have found four decoupled differential equations, where $V = K, K'$ represents the valley degrees of freedom and $i = A, B$ represents the sublattices in graphene.

Let us consider the differential equation (\ref{4.6}) and perform the following coordinate transformation $\xi = \frac{eB_0}{2} r^2$. This allows us to rewrite the differential equation as
\begin{eqnarray}
    \xi\frac{d^2\psi^{V}_{i}(\xi)}{d\xi^{2}} + \frac{d\psi^{V}_{i}(\xi)}{d\xi} + \left[\frac{\beta_i}{2eB_0} - \frac{\xi}{4} - \frac{(\mu^{V}_{i})^2}{4\xi}\right]\psi^{V}_{i}(\xi) = 0,
    \label{4.7}
\end{eqnarray}
where the parameters $\beta_i$ and $\mu^{V}_{i}$ are given by
\begin{eqnarray}
    \beta_i &=& E^2 - eB_0\left(\mu^V \pm \frac{\sigma_z}{2}\right) - \left(M +\nu\lambda_{SO}\sqrt{h^2 + p^2}\right)^2, \label{4.8} \\
    (\mu^{V}_{i})^2 &=& \left(\mu^V \mp \frac{\sigma_z}{2}\right)^2. \label{4.9} 
\end{eqnarray}
The asymptotic solution to equation (\ref{4.7}) can be write as follow
\begin{eqnarray}
    \psi^{V}_{i}(\xi) = e^{-\frac{\xi}{2}} \xi^{\frac{|\mu^{V}_{i}|}{2}} F(\xi),
    \label{4.10}
\end{eqnarray}
leading to
\begin{eqnarray}
    \xi\frac{d^2 F(\xi)}{d\xi^{2}} + (|\mu^{V}_{i}| + 1 - \xi)\frac{dF(\xi)}{d\xi} +  \left[\frac{\beta_i}{2e B_0} - \frac{1}{2} - \frac{|\mu^{V}_{i}|}{2}\right]F(\xi) = 0.
    \label{4.11}
\end{eqnarray}
This equation is the confluent hypergeometric differential equation \cite{Machado}, and therefore $F(\xi) =\ _{1}F_1(-n,|\mu^{V}_{i}|+1;\xi)$ is the confluent hypergeometric function. By truncating the confluent hypergeometric function, we can express it as an $n$-degree polynomial:
\begin{eqnarray}
    n = \frac{\beta_i}{2e B_0} - \frac{1}{2} - \frac{|\mu^{V}_{i}|}{2},
    \label{4.12}
\end{eqnarray}
with $n = 0, 1, 2, \ldots$ being an integer. Consequently, we can determine the energy levels for the disclinated graphene-based topological insulator as follows,
\begin{eqnarray}
    E = \pm\sqrt{2\omega\left[2n + \left|\mu^V -\tau_z \frac{\sigma_z}{2}\right| + \left(\mu^V -\tau_z \frac{\sigma_z}{2}\right) + 1 + \tau_z\sigma_z\right] + \left(M +\nu\lambda_{SO}\sqrt{h^2 + p^2}\right)^2}.
    \label{4.13}
\end{eqnarray}

\begin{figure}[t]
    \centering
    \begin{subfigure}{.5\textwidth}
        \centering
        \includegraphics[width=.8\linewidth]{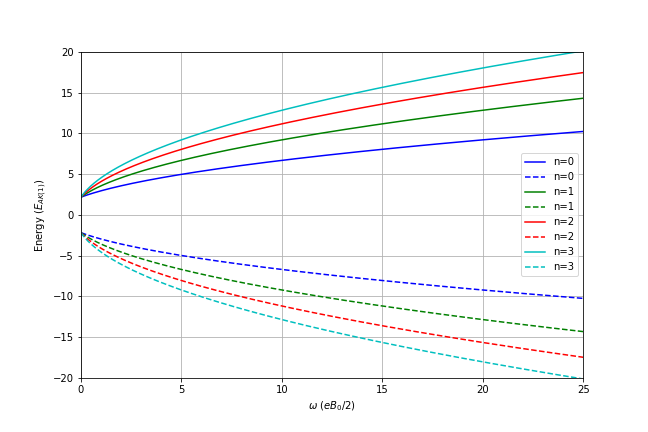}
        \caption{Sublattice $\mathcal{A}$ and $\nu = 1$}
    \end{subfigure}%
    \begin{subfigure}{.5\textwidth}
        \centering
        \includegraphics[width=.8\linewidth]{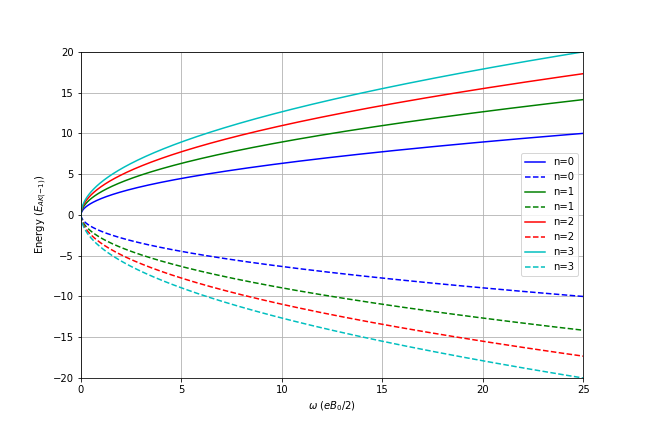}
        \caption{Sublattice $\mathcal{A}$ and $\nu = -1$}
    \end{subfigure}\\
    \begin{subfigure}{.5\textwidth}
        \centering
        \includegraphics[width=.8\linewidth]{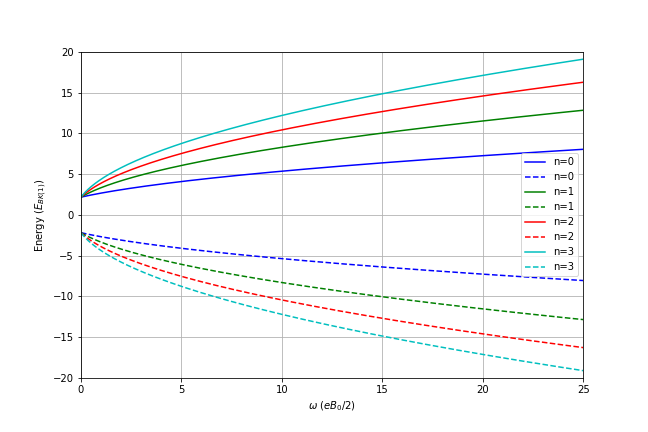}
        \caption{Sublattice $\mathcal{B}$ and $\nu = 1$}
    \end{subfigure}%
    \begin{subfigure}{.5\textwidth}
        \centering
        \includegraphics[width=.8\linewidth]{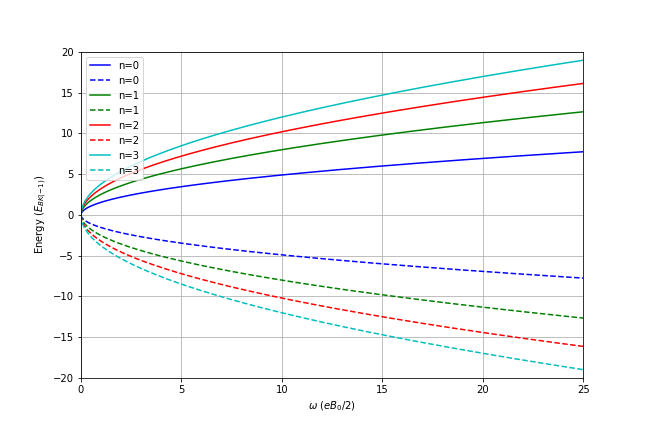}
        \caption{Sublattice $\mathcal{B}$ and $\nu = -1$}
    \end{subfigure}
    \caption{Landau levels for $K$ valley}
    \label{fig1}
\end{figure}

We can observe that the introduction of an external magnetic field gives rise to the Landau levels (\ref{4.13}), where $\omega = \frac{eB_{0}}{2}$ is the cyclotron frequency. In addition to the dependence on the external magnetic field, these Landau levels have the disclination influence implicit in the parameter $\mu^{V}$. Another important point is that the breaking of inversion symmetry introduces a gap in the energy spectrum through the $M$ mass term and the $\lambda_{SO}$ SOC term. Additionally, the bound states have their degeneracy split into four states due to the external magnetic field, and each of these four states is further split into two by the spin projection in the disclinated geometry, resulting in eight states. For specific values of the quantum flux $\Phi/\Phi_0$ and the mass term, we can obtain zero modes that are topologically protected. Therefore, disclinated graphene in the presence of an external magnetic field behaves as a topological insulator.

\begin{figure}[ht]
    \centering
    \begin{subfigure}{.5\textwidth}
        \centering
        \includegraphics[width=.8\linewidth]{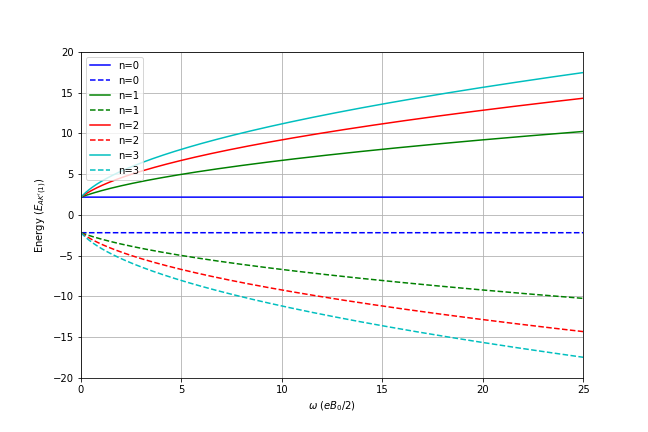}
        \caption{Sublattice $\mathcal{A}$ and $\nu = 1$}
    \end{subfigure}%
    \begin{subfigure}{.5\textwidth}
        \centering
        \includegraphics[width=.8\linewidth]{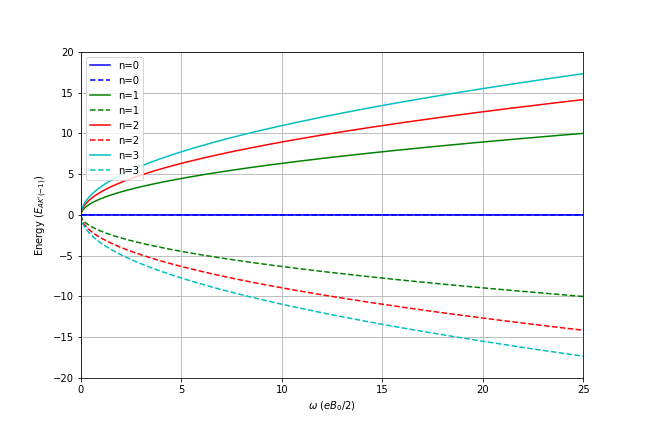}
        \caption{Sublattice $\mathcal{A}$ and $\nu = -1$}
    \end{subfigure}\\
    \begin{subfigure}{.5\textwidth}
        \centering
        \includegraphics[width=.8\linewidth]{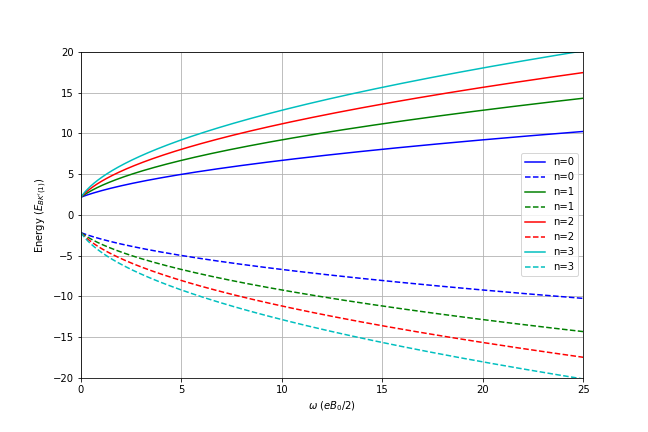}
        \caption{Sublattice $\mathcal{B}$ and $\nu = 1$}
    \end{subfigure}%
    \begin{subfigure}{.5\textwidth}
        \centering
        \includegraphics[width=.8\linewidth]{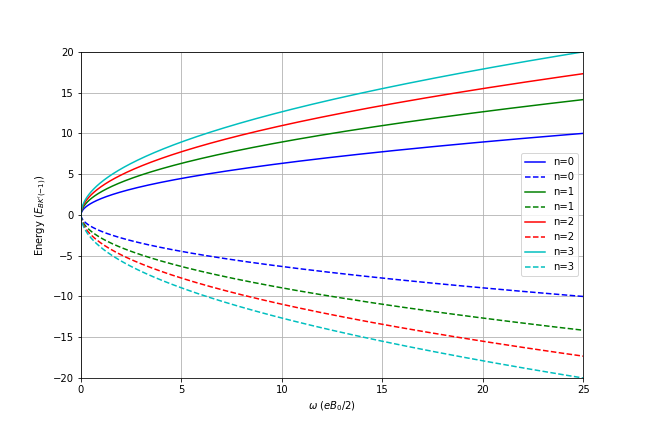}
        \caption{Sublattice $\mathcal{B}$ and $\nu = -1$}
    \end{subfigure}
    \caption{Landau levels for $K'$ valley}
    \label{fig2}
\end{figure}

In order to better visualize our conclusions about this problem, we will analyze the energy spectrum for different values of cyclotron frequency $\omega$. For this purpose, we choose $j=\frac{1}{2}$, $\frac{\Phi}{\Phi_0} = -\frac{2}{3}$, $N = 1$, and $M = \lambda_{SO}\sqrt{h^2 + p^2}$. We illustrate the Landau levels for the $K$ valley in Fig.~\ref{fig1} and for the $K'$ valley in Fig.~\ref{fig2}. For the $K$ valley (Fig.~\ref{fig1}), we observe that disclinated graphene behaves as an insulator, where the gap between the conduction and valence bands is controlled by the term $M + \nu\sqrt{h^2 + p^2}$. Therefore, we have a larger gap for $\nu = 1$ than for $\nu = -1$. For the $K'$ valley, the set of plots in Fig.~\ref{fig2} indicates that disclinated graphene remains an insulator, but now we can observe the presence of zero modes independent of the external magnetic field. In fact, for sublattice $\mathcal{A}$ and $\nu = -1$, we have topologically protected zero modes due to the breaking of inversion symmetry with a Rashba-like spin-orbit coupling. It is interesting to point out that the zero modes appear for only one component of the eight-component spinor and depend on the values of the quantum flux $\Phi/\Phi_0$. For example, we can obtain another zero mode in the $K$ valley, sublattice $\mathcal{B}$, and $\nu = -1$ for quantum flux $\frac{\Phi}{\Phi_0} = -\frac{7}{6}$. Another interesting consequence is the possibility of different gap width among sites, depending of spin-projector value $\nu$. Finally, we can write the spinor $\Psi$ with eight components as:
\begin{eqnarray}
    \Psi(\xi,\theta) = A_{j,\nu} e^{-\frac{\xi}{2}} e^{ij\theta} \left(\begin{array}{c}
    \xi^{\frac{|\mu^{K}_{A}|}{2}}F(-n, |\mu^{K}_{A}| + 1;\xi) \\ c_\nu \xi^{\frac{|\mu^{K}_{A}|}{2}} F(-n, |\mu^{K}_{A}| + 1;\xi) \\ \xi^{\frac{|\mu^{K}_{B}|}{2}} F(-n, |\mu^{K}_{B}| + 1;\xi)\\ c_\nu \xi^{\frac{|\mu^{K}_{B}|}{2}} F(-n, |\mu^{K}_{B}| + 1;\xi)\\ \xi^{\frac{|\mu^{K'}_{A}|}{2}} F(-n, |\mu^{K'}_{A}| + 1;\xi)\\ c_\nu \xi^{\frac{|\mu^{K'}_{A}|}{2}} F(-n, |\mu^{K'}_{A}| + 1;\xi)\\ \xi^{\frac{|\mu^{K'}_{B}|}{2}} F(-n, |\mu^{K'}_{B}| + 1;\xi)\\ c_\nu \xi^{\frac{|\mu^{K'}_{B}|}{2}} F(-n, |\mu^{K'}_{B}| + 1;\xi)
    \end{array}\right).
\end{eqnarray}

\section{Summary and conclusions}\label{sec5}

In this contribution, we have analyzed the disclinated massive graphene-based topological insulator in the presence of an external magnetic field. The massive graphene was described by the modified Kane-Mele Hamiltonian for the disclination geometry, where we obtain a Rashba-like SOC term. The introduction of the external magnetic field is responsible for the quasi-particle quantum dynamics. The energy spectrum is characterized by Landau levels defined by valleys, sublattices, and spin-projection degrees of freedom. The external magnetic field, along with the SOC and mass terms, breaks the wavefunction degeneracy, resulting in an eight-component spinor. The SOC and mass terms determine the zero modes and the gap width, and these zero modes are topologically protected by the breaking of the inversion symmetry. The zero modes occur for specific values of the defect quantum flux and appear in only one spinor component, and they are known as edge states. Another relevant point is the difference in gap length depending on the spin projection $\nu = \pm 1$, resulting in a gap difference among several sites in the graphene lattice.
  
{\bf Acknowledgements:} This work was supported by Conselho Nacional de Desenvolvimento Cient\'{\i}fico e Tecnol\'{o}gico (CNPq) and Funda\c{c}\~ao de Apoio a Pesquisa do Estado da Para\'iba (Fapesq-PB). G. Q. Garcia would like to thank to Fapesq-PB for financial support Grant BLD-ADT-A2377/2024. P. J. Porf\'irio would like to thank the Brazilian agency CNPq for financial support (PQ--2 grant, process No. 307628/2022-1). The work by C. Furtado has been supported by the CNPq project PQ Grant 1A No. 311781/2021-7 .

\end{document}